\documentclass[twocolumn]{aastex61}
\usepackage{graphicx}  
\usepackage{amsmath}

\defcitealias{2018RAA....18...52T}{Paper I}

\received{July 1, 2016}
\revised{September 27, 2016}
\accepted{\today}
\submitjournal{ApJS}

\shorttitle{LAMOST variable star candidates}
\shortauthors{Tian et al.}

\begin{document}

\title{A catalog of RV variable star candidates from LAMOST}

\correspondingauthor{Zhijia Tian; Xiaowei Liu}
\email{tianzhijia@ynu.edu.cn, x.liu@pku.edu.cn}


\author{Zhijia Tian}
\altaffiliation{LAMOST Fellow}
\affiliation{Department of Astronomy, Key Laboratory of Astroparticle Physics of Yunnan Province, Yunnan University, Kunming 650200, P. R. China}

\author{Xiaowei Liu}
\affiliation{South-Western Institute for Astronomy Research, Yunnan University, Chenggong District, Kunming 650500, P. R. China}

\author{Haibo Yuan}
\affiliation{Department of Astronomy, Beijing Normal University, Beijing 100875, P. R. China}

\author{Xuan Fang}
\affiliation{Key Laboratory of Optical Astronomy, National Astronomical Observatories of Chinese Academy of Sciences (NAOC), 20A Datun Road, Chaoyang District, Beijing 100101, China}
\affiliation{Department of Physics, Faculty of Science, The University of Hong Kong, Pokfulam Road, Hong Kong, P. R. China}

\author{Bingqiu Chen}
\affiliation{South-Western Institute for Astronomy Research, Yunnan University, Chenggong District, Kunming 650500, P. R. China}

\author{Maosheng Xiang}
\affiliation{Max-Planck Institute for Astronomy, Heidelberg, Germany}

\author{Yang Huang}
\affiliation{South-Western Institute for Astronomy Research, Yunnan University, Chenggong District, Kunming 650500, P. R. China}

\author{Shaolan Bi}
\affiliation{Department of Astronomy, Beijing Normal University, Beijing 100875, P. R. China}

\author{Wuming Yang}
\affiliation{Department of Astronomy, Beijing Normal University, Beijing 100875, P. R. China}

\author{Yaqian Wu}
\affiliation{National Astronomy Observatories, Chinese Academy of Sciences, Beijing 100012, P. R. China}

\author{Chun Wang}
\altaffiliation{LAMOST Fellow}
\affiliation{Department of Astronomy, Peking University, Beijing 100871, P. R. China}

\author{Huawei Zhang}
\affiliation{Department of Astronomy, Peking University, Beijing 100871, P. R. China}

\author{Zhiying Huo}
\affiliation{National Astronomy Observatories, Chinese Academy of Sciences, Beijing 100012, P. R. China}

\author{Yong Yang}
\affiliation{South-Western Institute for Astronomy Research, Yunnan University, Chenggong District, Kunming 650500, P. R. China}

\author{Gaochao Liu}
\affiliation{China Three Gorges University, Yichang 443002, P. R. China}

\author{Jincheng Guo}
\altaffiliation{LAMOST Fellow}
\affiliation{Department of Astronomy, Peking University, Beijing 100871, P. R. China}

\author{Meng Zhang}
\affiliation{Department of Astronomy, Peking University, Beijing 100871, P. R. China}



\begin{abstract}
RV variable stars are important in astrophysics. 
The Large Sky Area Multi-Object Fiber Spectroscopic Telescope (LAMOST) spectroscopic survey has provided $\sim$6.5 million stellar spectra in its Data Release 4 (DR4). 
During the survey $\sim$4.7 million unique sources were targeted and $\sim$1 million stars observed repeatedly. 
The probabilities of stars being RV variables are estimated by comparing the observed radial velocity variations with simulated ones. 
We build a catalog of 80,702 RV variable candidates with probability greater than 0.60 by analyzing the multi-epoch sources covered by the LAMOST DR4. 
Simulations and cross-identifications show that the purity of the catalog is higher than 80\%. 
The catalog consists of 77\% binary systems and 7\% pulsating stars as well as 16\% pollution by single stars. 
3,138 RV variables are classified through cross-identifications with published results in literatures. 
By using the 3,138 sources common to both LAMOST and a collection of published RV variable catalogs we are able to analyze LAMOST's RV variable detection rate. 
The efficiency of the method adopted in this work relies not only on the sampling frequency of observations but also periods and amplitudes of RV variables. 
With the progress of LAMOST, $\it{Gaia}$ and other surveys, more and more RV variables will be confirmed and classified. 
This catalog is valuable for other large-scale surveys, especially for RV variable searches. 
The catalog will be released according to the LAMOST Data Policy via http://dr4.lamost.org.
\end{abstract}






\section{Introduction} \label{sec:intro}
Binary stars play a crucial role in astrophysics. 
Statistics and identifications of binary systems are significant for several reasons, the major ones being that such basic issues as star formation and evolution, the initial mass function (IMF) and Galactic chemical evolution are all influenced by the binary properties of the stellar population. 
Despite the high fraction of binary stars ($\sim$ 50\% for main-sequence stars), our understandings of the physics of binary stars are still at a basic stage. 
\citet[][]{2010ApJS..190....1R} presents the multiplicity of 454 solar-type stars within 25 pc at high completeness. They show that early-type and metal-poor stars dominate higher binary factions than late-type and metal-rich stars. 
The period distribution of the sample follows a log-normal distribution with a median of about 300 years. 
Meanwhile, early- and late-type stars do not stem from the same parent period distribution \citep{2011A&A...529A..92K}. 
The discrepancy is yet to be explained and could be related to the mechanism of binary formation.

A summary on empirical knowledge of stellar multiplicity for embedded protostars, pre-main-sequence, main sequence, and brown dwarfs is performed by \citet{2013ARA&A..51..269D}. It is demonstrated that the multiplicity rate and breadth of the orbital period distribution are steep functions of the primary mass and environment.
More efforts in recent years have been made in analyses of binary fractions based on large samples of survey data \citep[e.g.][hereafter \citetalias{2018RAA....18...52T}]{1991A&A...248..485D,2014ApJ...788L..37G,2017MNRAS.469L..68G,2015ApJ...799..135Y,2018ApJ...854..147B,2018RAA....18...52T}. 
These works investigate the binary fractions against stellar parameters, i.e. mass, $T_{\rm eff}$, and abundance. 
All the researches indicate that metal-poor stars have a higher binary fraction than metal-rich stars.  
However, metal-rich disk stars are found to be 30\% more likely to have companions with periods shorter than 12 days than metal-poor halo stars \citep{2015ApJ...806L...2H}. 
The binary fraction is not only related to stellar parameters but also orbital periods \citep{2001MNRAS.326.1391M,2017ApJS..230...15M}. 

Besides estimating binary fractions in large samples, identifications of binary systems have been carried out. 
The American Association of Variable Star Observers (AAVSO) contributes to building an International Variable Star Index \citep[VSX;][]{2006SASS...25...47W}. 
A database of thousands of eclipsing binaries is established \citep[][and references therein]{Mat12} with $\it Kepler$ light curves \citep{2010Sci...327..977B,2010ApJ...713L..79K}. 
\citet{2014ApJS..213....9D} presents $\sim$47,000 periodic variables found during the analysis of 5.4 million variable star candidates covered by the Catalina Surveys Data Release-1 \citep[CSDR1,][]{2012AAS...21942820D}, and investigates the rate of confusion between objects classified as contact binaries and type c RR Lyrae (RRc's) based on periods, amplitudes, radial velocities and stellar parameters. 
The General Catalog of Variable Stars (GCVS) containing binary stars is released in the latest version \citep[GCVS Version 5.1,][]{2017ARep...61...80S}.  
The Binary star DataBase (BDB) collects data on physical and positional parameters of 240,000 components of 110,000 multiple-star systems \citep{2015A&C....11..119K}. 
\citet{2018AJ....156...18P} makes use of the multi-epoch data obtained with the APOGEE \citep{2017AJ....154...94M,2018ApJS..235...42A} and selects $\sim$ 5000 evolved stars with probable companions. 
To build a sample of distant halo wide binaries, \citet{2018MNRAS.480.4302C} searches stellar pairs with small differences in proper motion and small projected separation on the sky as binary candidates, and validates the sample through RVs from medium and low-resolution spectra obtained with SDSS \citep{2000AJ....120.1579Y}. 
Binaries and Triples are identified using high-dispersion spectra, which can be much better fit with a superposition of two or three model spectra, drawn from the same isochrone, than any single-star model.
 \citet{2018MNRAS.476..528E} applies the data-driven spectral model to APOGEE DR13 spectra of main-sequence stars and identifies unresolved multiple-star systems. 
$\it{Gaia}$ Data Release 2  \citep[$\it{Gaia}$ DR2, ][]{2018A&A...616A...1G} enables catalogs of variable stars \citep{2019A&A...622A..60C,2018A&A...618A..58M,2018A&A...620A.197R,2019A&A...625A..97R}. 

However, binary identification based on RVs derived from a low-dispersion spectroscopic spectra survey is still almost blank. 
Fortunately, Large Sky Area Multi-Object Fiber Spectroscopic Telescope (LAMOST) provided millions of stellar spectra, of which about 20\% of the targets have been observed repeatedly. 
The quantity of these spectra can enhance time-domain studies of stars, stellar parameters, and their RVs and help select and confirm variable candidates. 
We build a catalog of RV variable candidates detected with the LAMOST. 
In Section 2, we describe the data used in this work. The method is presented in Section 3. 
The results are shown in Section 4, followed by discussions and conclusions in Section 5.

\section{Data} \label{sec:data}
\begin{figure}[ht!]
\centering\includegraphics[width=0.90\columnwidth]{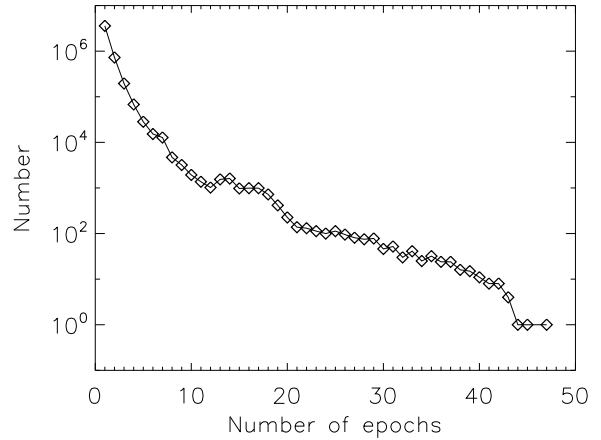}
\caption{The number of stars against number of epochs in LAMOST DR4.\label{fig:general}}
\end{figure}

The LAMOST spectroscopic survey provides the largest database of low-resolution ($\mathit{R}$$\sim$2000) spectra to measure stellar atmospheric parameters and radial velocities for millions of stars \citep{2012RAA....12.1197C,2012RAA....12..723Z,2012RAA....12..735D,2014IAUS..298..310L,2015MNRAS.448..855Y}. 
The survey has obtained $\sim$6.5 million spectra for $\sim$4.7 million unique stars in its DR4. 
In this survey, $\sim$1 million stars have been observed in 2 to over 40 epochs. 
As presented in Fig. \ref{fig:general}, the number of stars decreases with the number of epochs nearly exponentially. 
Through comparing multi-epoch observations especially for RVs, variable stars could be detected. 
We adopt the RV and stellar parameters yielded by LAMOST stellar parameter pipeline at Peking University $-$ LSP3 \footnote{\url{http://dr4.lamost.org/v2/doc/vac}} \citep{2015MNRAS.448..822X,2017MNRAS.464.3657X} to select variable stars. 
The pipeline estimates RV through cross-correlating with an ELODIE template \citep[][]{2001A&A...369.1048P,2007astro.ph..3658P} with close values of atmospheric parameters.  
For the determinations of stellar atmospheric parameters (e.g. $T_{\rm eff}$, $\log g$ and [Fe/H]), templates from the MILES library \citep{2006MNRAS.371..703S,2011A&A...532A..95F}, obtained with a spectral resolving power similar to that of LAMOST spectra and accurately flux calibrated, are used instead. 
As discussed in \citet{2015MNRAS.448..822X}, the MILES spectra with low-resolution are wavelength calibrated to an accuracy of only approximately 10 $\rm km \ s^{-1}$, not good enough for the purpose of RV determinations for the LAMOST spectra. However, the ELODIE library of high-resolution spectra is much proper to be used as RV templates.  
Furthermore, the LSP3 estimates RV prior to atmospheric parameters, which avoids systemic  uncertainties of RV caused by adopting different spectral libraries in the pipeline. 
In the latest version of LSP3, 267 new template spectra obtained using the NAOC 2.16-m telescope and the Yunnan Astronomical Observatory (YAO) 2.4-m telescope \citep[obtained by][]{2018MNRAS.480.4766W} have been added to the MILES library to generate parameter estimates \citep{2017MNRAS.464.3657X}. 

The LSP3 pipeline ignores the effects of binary stars when estimating RV and other stellar parameters.
Most of the stars have a radial velocity error of a few $\rm km\ s^{-1}$. However, some of them, mostly hot stars with low signal-to-noise ratios (SNRs), have errors as large as 20 $\rm km \ s^{-1}$ \citep{2015MNRAS.448..822X,2017MNRAS.464.3657X}. To identify binary systems or candidates reliably, we limit the SNR of spectra greater than 10. 

\subsection{RVs and their uncertainties}
\begin{figure}[ht!]
\centering\includegraphics[width=0.85\columnwidth]{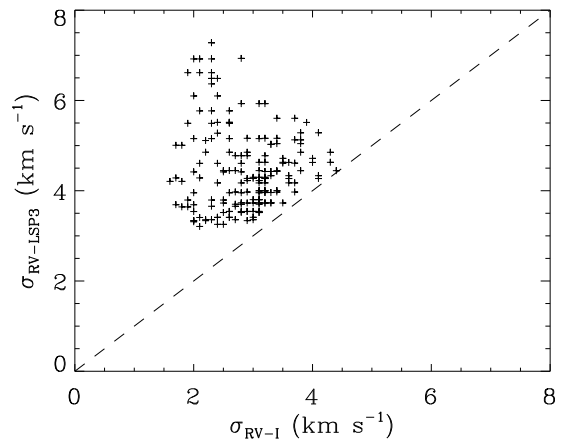}
\caption{The comparison of $\sigma_{\rm RV}$ from LSP3 and \citetalias{2018RAA....18...52T} .\label{fig:rvcom}}
\end{figure}

As discussed in \citet{2015MNRAS.448..822X,2017MNRAS.464.3657X}, the $\sigma_{\rm RV}$ is quite sensitive to SNR and depends on other stellar parameters.
The LSP3 pipeline estimates $\sigma_{\rm RV}$ by comparing RVs from multi-epoch observations of similar SNRs and
spectral types, assuming that $\sigma_{\rm RV}$ is contributed from random error following a Gaussian distribution and systematic error. 
It considers the stars as single ones and ignores the influence of binary stars on RV, 
and attributes the variation in RV as uncertainties and therefore over-estimates $\sigma_{\rm RV}$. 
The $\sigma_{\rm RV}$ has been reappraised in \citetalias{2018RAA....18...52T} when estimating the binary fraction ($\textit{f}_{\rm B}$) of dwarfs with SNR $>$ 50, taking into account the degeneracy between $\textit{f}_{\rm B}$ and $\sigma _{\rm RV}$. 
A comparison of the RV uncertainties from LSP3 (${\sigma _{\rm RV}}_{-\rm LSP3}$) and those from \citetalias{2018RAA....18...52T}  (${\sigma _{\rm RV}}_{-\rm I}$) is presented in Fig.  \ref{fig:rvcom}, which shows that the LSP3 pipeline over-estimates the uncertainties of RVs. 
The median $\sigma_{\rm RV}$ of dwarfs with SNR $>$ 50 is around 2.9 $\rm km \ s^{-1}$ while for the LSP3 pipeline it is $\sim$1.5 times higher at 4.3 $\rm km \ s^{-1}$. 
The precision of RVs with high SNR is adequate enough to detect short-period binaries. 

\begin{figure}[ht!]
\centering\includegraphics[width=0.85\columnwidth]{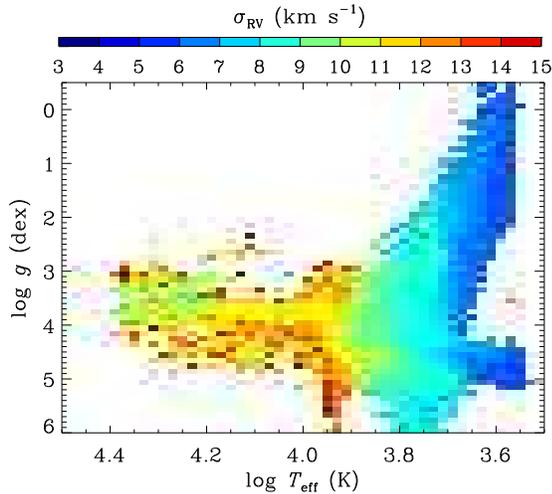}
\caption{The mean $\sigma_{\rm RV}$ for stars in the Hess diagram. \label{fig:eerr}}
\end{figure}

\begin{figure}[ht!]
\centering\includegraphics[width=0.85\columnwidth]{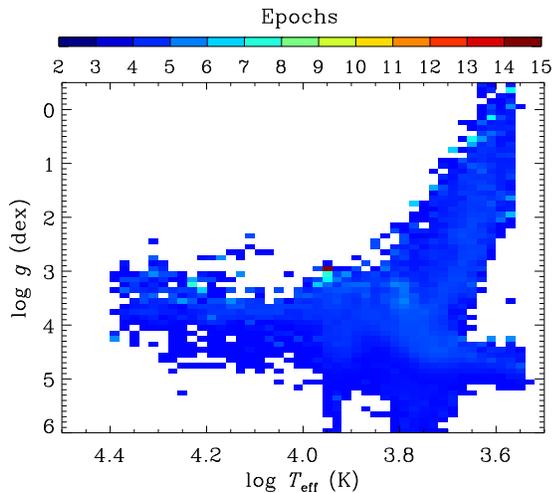}
\caption{The mean number of epochs for stars in the Hess diagram. \label{fig:hrepoch}}
\end{figure}

Figure \ref{fig:eerr} presents the distribution of mean $\sigma _{\rm RV}$ in the Hess diagram, which shows that $\sigma _{\rm RV}$ of hot stars are higher than those of cooler stars.  
The distribution of the average number of epochs in the Hess diagram is shown in Fig. \ref{fig:hrepoch} .
The distribution of the multi-epoch observations are uniform, which indicates that the $\sigma_{\rm RV}$ are not biased by selection effects of epochs. 

\subsection{Reliability of the data}
In this work, we adopt the RVs and $\sigma_{\rm RV}$ from LSP3 in our binary identification. 
For a single star with multiple observations in the same condition, the RVs obey a Gaussian distribution with a mean $\overline{\rm RV}$ 
and variance $\sigma_{\rm RV}^2$. 
However, for a star observed repeatedly in different conditions, we have a sample of $\rm RVs$ for the star, $\rm RV_{1}$, $\rm RV_{2}$, . . ., $\rm RV_{n}$, where each $\rm RV$ value is from a Gaussian distribution having the same mean $\overline{\rm RV}$ but a different standard deviation $\sigma_{\rm RV_{i}}$. 
The weighting factor is the inverse of $\sigma_{\rm RV_{i}}^2$, 
thus, the weighted radial velocity $\overline{\rm RV}$ is expressed as \footnote{\url{https://ned.ipac.caltech.edu/level5/Leo/Stats4_5.html}} 
\begin{equation}
\label{equ:rv}
\overline{\rm RV} =  \frac{\sum_{i=1}^{n}\rm RV_{i}/\sigma_{\rm RV_{i}}^2}{\sum_{i=1}^{n}1/\sigma_{\rm RV_{i}}^2},
\end{equation}
where the error of the $\overline{\rm RV}$ is 
\begin{equation}
\label{equ:s2}
\hat{\sigma}_{\overline{\rm RV}} =  \sqrt{\frac{1}{\sum_{i=1}^{n}1/\sigma_{\rm RV_{i}}^2}},
\end{equation}
and the weighted error is 
\begin{equation}
\label{equ:rverr}
\overline{\sigma_{\rm RV} } =  \frac{\sum_{i=1}^{n}\sigma_{\rm RV_{i}}/\sigma_{\rm RV_{i}}^2}{\sum_{i=1}^{n}1/\sigma_{\rm RV_{i}}^2}.
\end{equation}
The variance of the RV is 
\begin{equation}
\label{equ:rverr4}
S^2 = \frac{1}{n} \sum_{i=1}^{n} \frac{ (\rm RV_{i}- \overline{\rm RV})^2}{\sigma_{\rm RV_{i}}^2}.
\end{equation}

\begin{figure}[ht!]
\centering\includegraphics[width=0.9\columnwidth]{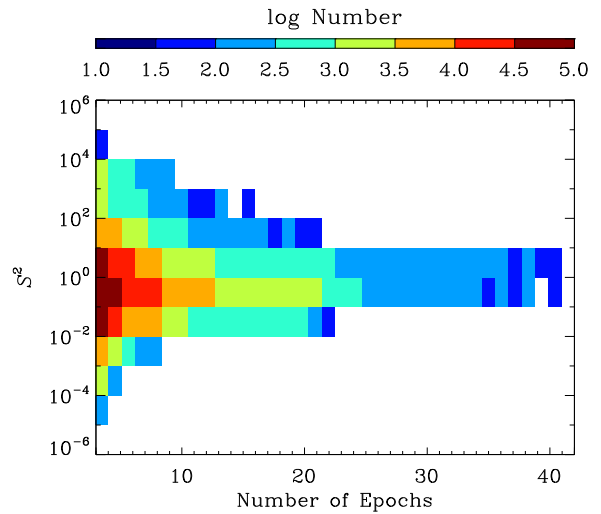}
\caption{The distribution of the multi-epoch sources against the $S^2$ and number of epochs. \label{fig:epchis}}
\end{figure}

The distribution of $S^2$ for the sources with multiple epochs is shown in Fig. \ref{fig:epchis}. The $S^2$ converges into 1 with enough epochs, which proves the validity of RVs with errors. 
Although the $\rm RVs$ of a binary or other RV variable star don't follow a normal distribution, we could also define their $\overline{\rm RV}$ and ${\sigma}_{\rm RV}$ through equations \ref{equ:rv} and \ref{equ:rverr}. 

\section{Method}\label{sec:method}
 \subsection{Feasibility analysis} \label{subsec:fa}

\begin{figure}[hbp]
\centering\includegraphics[width=0.9\columnwidth]{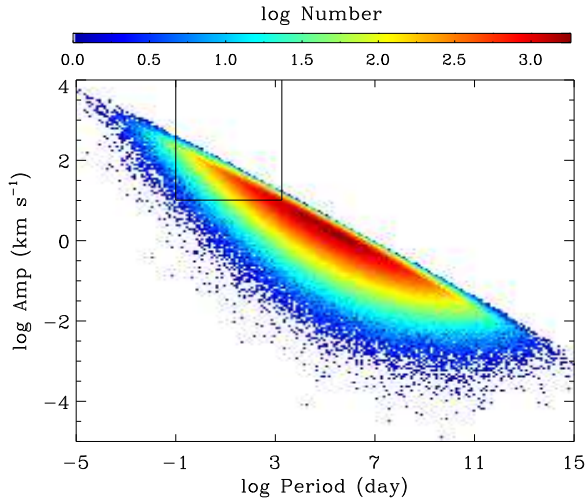}
\caption{The joint distribution of periods and amplitudes for the mock binaries. The box marks out the detection limit based on the LAMOST's capability. \label{fig:apdis}}
\end{figure}

In order to analyze the feasibility of detecting binaries through $\Delta \rm RV_{\rm max}$, a simulation is performed. We construct a sample of 1 million binary stars and count the percentage of stars detected based on the LAMOST's capability. 
For the binary systems $M_{B}$, we assume that: (1) the RVs are contributed by their primary stars; 
(2) their orbital orientations are isotropic in 3D space and initial phases follow a uniform distribution; 
(3) their primary masses follow the measured mass distribution of the LAMOST sample, which are determined by fitting the atmospheric parameters with the Yonsei-Yale (YY) isochrones \citep[][and references therein]{2004ApJS..155..667D}; 
(4) the mass ratio $q$ follows a power-law distribution \citep[$f(q) \propto q^{0.3 \pm 0.1}$, e.g.][]{2013ARA&A..51..269D}; 
(5) for the orbital period distribution, a log-normal profile \citep[with a mean value of $\log P$ = 5.03 and a dispersion of $\sigma_{\log P}$ = 2.28, where $P$ is in units of days, see][]{2010ApJS..190....1R} is adopted. 
The $\sigma_{\rm RVs}$ adopted in the simulation follow those derived from the LAMOST DR4 data. 
As shown in Fig. \ref{fig:apdis}, the amplitudes of the simulated binary stars are strongly dependent on the period distribution. 
Considering that the typical exposure time of each observation is about one hour and the time span of LAMOST DR4 is less than 5 years, the detection is more efficient for binary systems with periods in the range of 0.1 day to 5 years rather than those with extremely short or long periods. 
We adopt 10 $\rm km \ s^{-1}$ ($\sim$ 3.0$\sigma_{\rm RV}$ for dwarfs with SNR higher than 50) as a threshold of RV amplitude to recognize RV variable stars in the simulation. 
The box in the figure marks out the 12\% of simulated binaries detectable with LAMOST based on these thresholds. 
It demonstrates that a certain proportion of binary stars are detectable based on the LAMOST observations. 

\subsection{Probability of belonging to a binary system}
\label{method1}

\begin{figure}[ht!]
\centering\includegraphics[width=0.95\columnwidth]{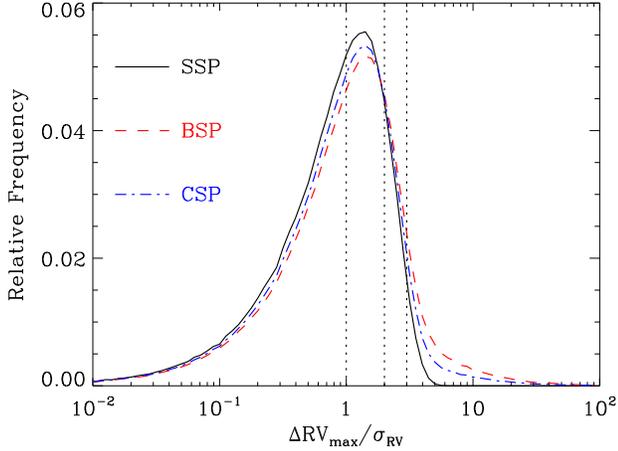}
\caption{The histograms of $\Delta \rm RV_{\rm max}/ \sigma_{RV}$ for the SSP, BSP and CSP samples, respectively. 
The vertical dotted-lines denote the thresholds of $\Delta \rm RV_{\rm max}/ \sigma_{RV}$ equal to 1, 2 and 3, respectively. \label{fig:simu3}}
\end{figure}

\begin{figure}[ht!]
\centering\includegraphics[width=0.95\columnwidth]{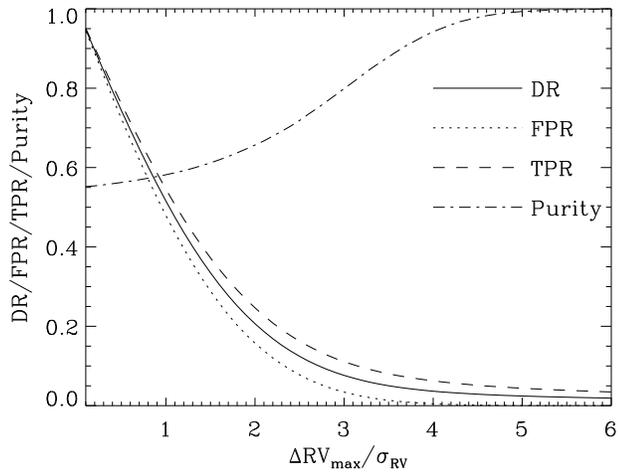}
\caption{The detection rate (DR), false positive rate (FPR), true positive rate (TPR) and purity against cutoff of $\Delta \rm RV_{\rm max} / \sigma_{\rm RV}$ for the CSP are plotted with solid, dotted, dashed and dash-dotted lines, respectively. \label{fig:fap}}
\end{figure}

The binary system could be identified by comparing $\Delta \rm RV_{\rm max}$ with ${\sigma_{\rm RV}}$, where the $\Delta \rm RV_{\rm max}$ presents the maximum radial velocity difference between any two epochs for the same object \citep[e.g.][]{2012ApJ...751..143M}. 
In order to test the effectiveness of the method, we mock three samples and count the percentages of detected stars at different thresholds. 
The three samples are defined as: 
\begin{enumerate}
\item [a.] Single Stellar Population (SSP) sample; 
\item [b.] Binary Stellar Population (BSP) sample; 
\item [c.] Composite Stellar Population (CSP, composed of $45\%$ single and $55\%$ binary systems) sample.
\end{enumerate}
The assumptions for the simulated samples are the same as those described in Section \ref{subsec:fa}.
The time separations of the multi-epoch observations are derived from the LAMOST DR4 data. 
The binary fraction $55\%$ adopted in the CSP is the median value derived from the LAMOST \citep[][]{2015ApJ...799..135Y,2014ApJ...788L..37G,2018RAA....18...52T}. 
Each sample consists of 1 million stars or systems. 
Note that intrinsic variables, e.g. pulsating stars, are ignored in these simulations. 
Under these assumptions,  the distributions of $\Delta \rm RV_{\rm max}$ for the SSP, BSP and CSP samples are constructed and presented in Fig. \ref{fig:simu3}. 
The vertical dashed-lines from left to right in the figure mark the cutoffs of  $\Delta \rm RV_{\rm max}/ \sigma_{RV}$ equal to 1, 2 and 3, respectively. 
The BSP sample has less low-value $\Delta \rm RV_{\rm max}$ and more high-value $\Delta \rm RV_{\rm max}$ than the SSP sample. 
The low-value $\Delta \rm RV_{\rm max}$ are dominated by random errors, while the high-value $\Delta \rm RV_{\rm max}$ are produced by variations of binary phases in the BSP (and the CSP). 
The detection rate (DR), false positive rate (FPR), true positive rate (TPR) and the fraction of real RV variables (purity) of the identified binaries against cutoffs of  $\Delta \rm RV_{\rm max}/ \sigma_{RV}$  are presented in Fig. \ref{fig:fap}. 
Improving the threshold of $\Delta \rm RV_{\rm max}/ \sigma_{RV}$ will increase the purity of the catalog, but reduces the DR at the same time.
Here the threshold of $\Delta \rm RV_{\rm max} > 3.0\sigma_{RV}$ is adopted to identify RV variable stars.
There are $3\%$, $11\%$ and $8\%$ stars with $\Delta \rm RV_{\rm max}$ greater than 3.0$\sigma_{\rm RV} $ in the SSP, BSP and CSP samples, respectively.  
The stars with $\Delta \rm RV_{\rm max}/ \sigma_{RV} > 3.0$ in the CSP consist of $20\%$ single stars and $80\%$ binary systems.
It indicates that the RV variable stars detected with the following method may be polluted by single stars. 

\begin{deluxetable*}{ccccc}[htp]
\tablecaption{The detection rate (DR), false positive rate (FPR), true positive rate (TPR) and purity for the CSP when adopting different cutoffs of $\Delta \rm RV_{\rm max}/ \sigma_{RV}$. \label{tab:fap}}
\tablecolumns{5}
\tablewidth{0pt}
\tablehead{
Rate & $\Delta\rm RV_{\rm max} > 1.0 \sigma_{\rm RV}$ & $\Delta\rm RV_{\rm max} > 2.0\sigma_{\rm RV}$ & $\Delta\rm RV_{\rm max} > 3.0\sigma_{\rm RV}$ & $\Delta\rm RV_{\rm max} >  4.0\sigma_{\rm RV}$}
\startdata
DR     & 0.515 & 0.206 & 0.076 & 0.036 \\
FPR   & 0.479 & 0.157 & 0.034 & 0.005 \\
TPR   & 0.545 & 0.246 & 0.110 & 0.062 \\
purity & 0.582 & 0.656 & 0.798 & 0.942 \\	
\enddata
\end{deluxetable*} 

\begin{figure}[ht]
\centering\includegraphics[width=0.95\columnwidth]{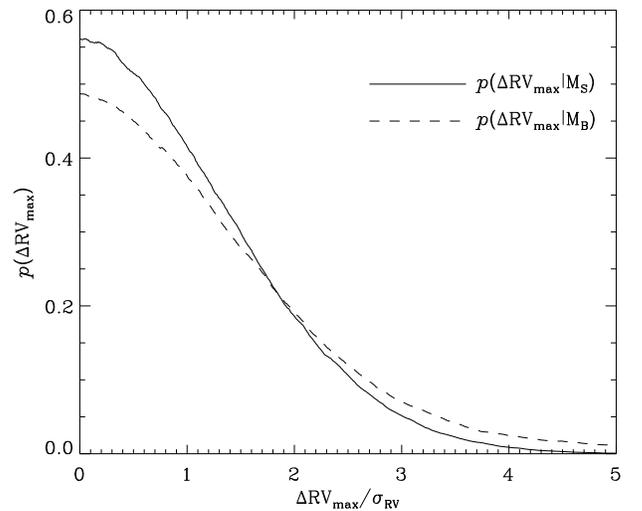}
\caption{The probability of obtaining $\Delta \rm RV_{\rm max}$ based on the single (SSP) and binary (BSP) assumptions, respectively.\label{fig:pv}}
\end{figure}

\begin{figure}[ht]
\centering\includegraphics[width=0.95\columnwidth]{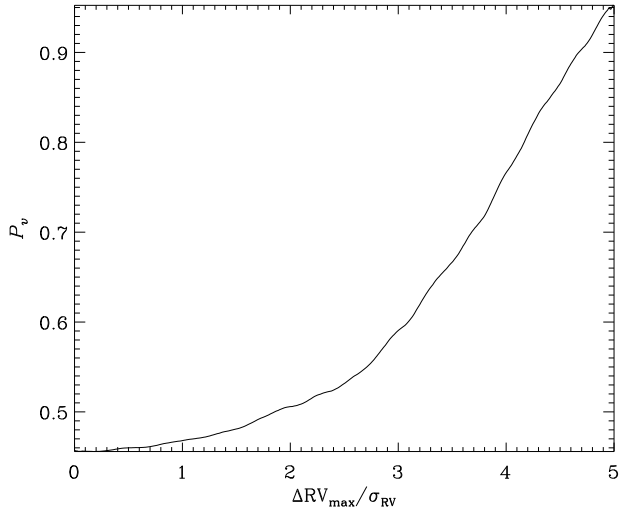}
\caption{The estimated probability $P_{v}$ of being a binary system based on the CSP with a binary fraction of $55\%$. \label{fig:pv0}}
\end{figure}

Given the value of $\Delta \rm RV_{\rm max}$ from observations, the probability of the star being a binary could be calculated based on the CSP simulation using Bayes' theorem:
\begin{equation}
\label{equ:pb}
\begin{split}
P_{v} & = p(M_{B}|\Delta \rm RV_{\rm max}) =  \frac{\it p(M_{B} \rm \Delta \rm RV_{\rm max}) }{\it p \rm (\Delta \rm RV_{\rm max}) }  \\
& =  \frac{\it p \rm (\Delta \rm RV_{\rm max} | \it M_{B} ) \  \it p \it (M_{B})  }{\it p \rm (\Delta \rm RV_{\rm max} | \it M_{B} ) \ \it p(M_{B})  + \it p \rm (\Delta \rm RV_{\rm max} | \it M_{S} ) \ \it p \rm ( \it M_{S})  } ,
\end{split}
\end{equation}
where $p(M_{B})$ and $p(M_{S})$ denote prior binary and single star fractions, respectively. 
Here we adopt a $p(M_{B})$ of 55\% derived from the LAMOST.
The $p(\Delta \rm RV_{\rm max} | \it M_{B}) $ and $p(\Delta \rm RV | \it M_{S})$ indicate the probabilities of obtaining $\Delta \rm RV_{\rm max}$ based on assumptions of the BSP and SSP models, respectively. 
Their values as functions of $\Delta \rm RV_{\rm max}/ \sigma_{RV}$ are shown in Fig. \ref{fig:pv}. 
For stars with $\Delta \rm RV/ \sigma_{RV} < 1.9$, they are more likely to be a single star rather than a binary system. The probability of being a binary system $P_{v}$ as a function of $\Delta \rm RV_{\rm max}/\sigma_{\rm RV}$ calculated through equation \ref{equ:pb} is presented in Fig. \ref{fig:pv0}. 
The higher value of $\Delta \rm RV_{\rm max}/ \sigma_{\rm RV}$ is, the higher probability of the star belonging to a binary system. 

This method is more sensitive to short-period binary stars, since their RVs vary more rapidly than long-period ones. 
For long-period (e.g. $\sim$ 300 years) binary stars, the time span of the LAMOST DR4 observations ($\sim$5 years) is too short to produce a large $\Delta \rm RV_{\rm max}$ to test their binarity efficiently. 

\begin{figure}[ht]
\centering\includegraphics[width=0.95\columnwidth]{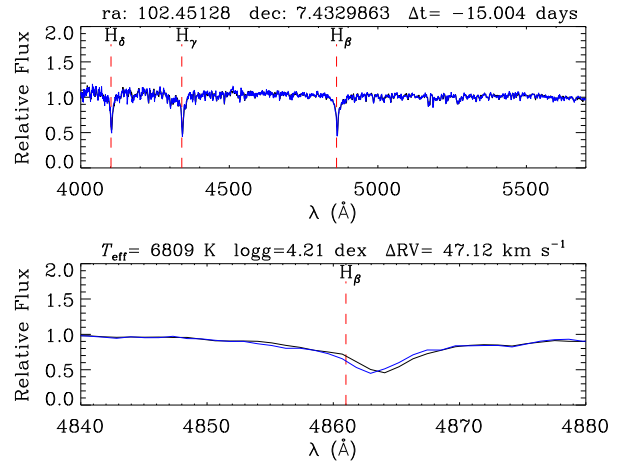}
\caption{Normalized spectra from two epochs (black and blue lines) for a representative star. The rest wavelengths of Balmer lines are plotted with vertical dashed lines. 
\label{fig:spec}}
\end{figure}

The Balmer lines are covered in the blue arm (3700 - 5900$\rm \AA$) of LAMOST.  
Figure \ref{fig:spec} plots the normalized LAMOST spectra for a representative star at two different epochs. The shift of $\rm H_{\beta}$ is clearly seen in the bottom panel, demonstrating LAMOST's capability to measure $\Delta \rm RV_{\rm max}$. 
Here we measure the depths of $\rm H_{\beta}$ from the normalized spectra. In order to ensure the reliability of RV measurements, we eliminate the sources with $\rm H_{\beta}$ depths less than 0.3. 
Meanwhile, the sources with high $\Delta \rm RV_{\rm max}$ values are confirmed by visual inspections to identify and remove the spectra affected by cosmic rays. 

\section{Catalog of RV variable stars}
\label{sec:result}
We apply the method to the LAMOST (DR4) data and estimate the binary probabilities of stars.  
Here we adopt a threshold of $P_{v} >$ 0.6 ($\Delta \rm RV_{\rm max} > 3.0\sigma_{RV}$) to identify binary stars and build a catalog of binary candidates. 
According to the simulation of CSP in Section \ref{method1}, the FPR is about 3\% at this threshold based on the capability of LAMOST. 
Since the cumulative run time of LAMOST is much less than the mean period of binary systems,  
 the LAMOST data is not suitable for detecting long-period binaries. 
There are $\sim$120,000 stars with $P_{v} >$ 0.6 ($\Delta \rm RV_{\rm max} > 3.0\sigma_{RV}$) in the LAMOST's DR4 sources with multiple epochs.  
After adopting the criteria of spectral depth and visual inspections, an assemblage of 80,702 RV variable star candidates remains in our final catalog as listed in Table \ref{tab:cat}. 
Note that in the simulation we only consider single and binary stars, but the sample observed with the LAMOST includes some intrinsic variables such as pulsating stars. 

\begin{figure}[!hpb]
\centering\includegraphics[width=0.9\columnwidth]{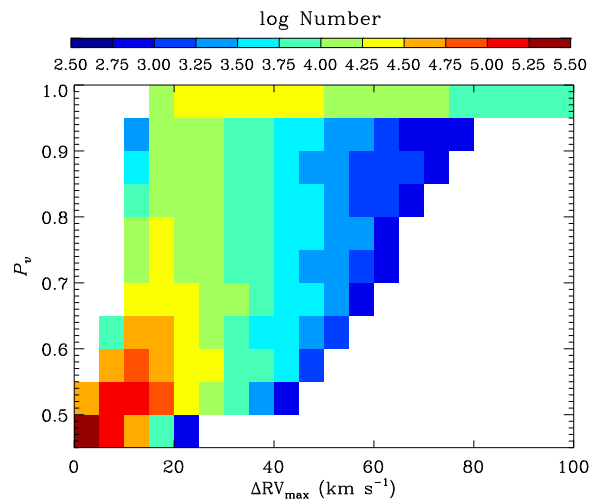}
\caption{The distributions of stars against $\Delta \rm RV_{max}$  and $P_{v}$.\label{fig:drvpv}}
\end{figure}

\begin{figure}
\centering\includegraphics[width=0.9\columnwidth]{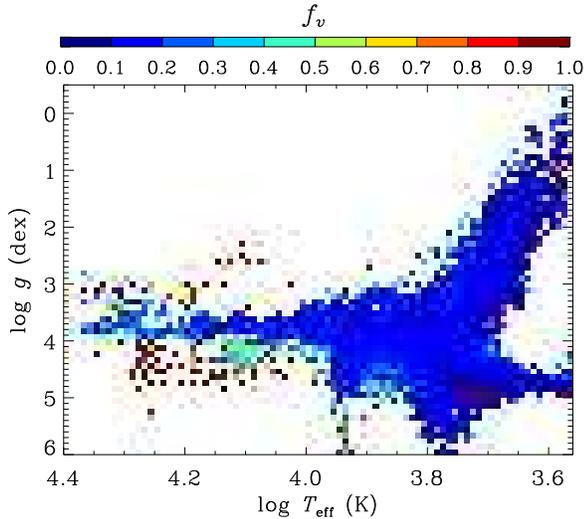} 
\caption{The fractions $f_{v}$ of stars with $P_{v} >$ 0.6 in each bin of Hess diagram.\label{fig:ssll3}}
\end{figure}

The distribution of the repeatedly observed stars in two-dimensional space of $\Delta \rm RV_{max}$ versus $P_{v}$ is shown in Fig. \ref{fig:drvpv}. 
The majority of the repeated targets that dominate low $P_{v}$ values ($P_{v} <$ 0.6) are single stars or unrecognized RV variables. 
Meanwhile, we present the fraction $f_{v}$ of stars with  $P_{v} >$ 0.6 in each bin with a size of 0.02 dex by 0.2 dex for $\log T_{\rm eff}$ and $\log g$ respectively in Fig. \ref{fig:ssll3}. As shown in the figure, the extended distribution of main-sequence stars with $P_{v} >$ 0.6 is broader than those with $P_{v} <$ 0.6. 
Stars with high $P_{v}$ have higher probabilities of being binaries than those with low $P_{v}$. 

\subsection{The purity of the catalog}
In order to verify the purity of the catalog and estimate pollutions by single stars, we perform a cross-identification between the LAMOST multi-epoch sources and a catalog of RV standard stars  published by \citet{2018AJ....156...90H} based on the APOGEE data \citep{2017AJ....154...94M,2018ApJS..235...42A}. 
There are 1,274 common sources between them. One hundred and three RV standard stars among the common sources have $P_{v} >$ 0.6. 
It means a single star contribution $\sim 8\%$ to our catalog. The purity of our catalog is approximately 92\%, which agrees with the simulation in Section \ref{method1}. 
Considering the cross-identification between our catalog and \citet{2018AJ....156...90H}, as well as the pollution by single stars in the simulation from Section \ref{method1}, the purity of our catalog is estimated to be higher than $\sim 80\%$. 

\subsection{Cross-match with $\it Kepler$ Eclipsing Binaries}\label{subsec:keb}
\begin{figure}
\centering\includegraphics[width=0.9\columnwidth]{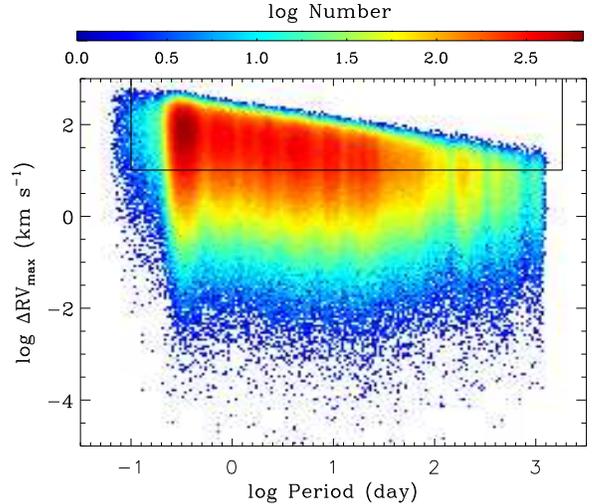}
\caption{The joint distribution of periods and $\Delta \rm RV_{\rm max}$ for the mock eclipsing binaries. The box marks out the detection limit based on the LAMOST's capability. }\label{fig:apdis_keb}
\end{figure}

A database of thousands of  $\it Kepler$ eclipsing binaries (KEBs) is released by \citet[][and references therein]{Mat12}. 
In total 520 KEBs have been observed repeatedly by the LAMOST-$\it Kepler$ project that uses the LAMOST to make spectroscopic follow-up observations for the $\it Kepler$ targets \citep{2015ApJS..220...19D,2018ApJS..238...30Z}. 
Of those, 255 stars are detected as binary stars in our catalog based on the LAMOST observations. 
To test the rationality of such application on the $\it Kepler$ data, we simulate a sample of 1 million eclipsing stars and count the rate of the detectable binaries. The assumptions of the mock sample are similar to those described in Section \ref{method1}. However, for the simulated eclipsing stars, we fix the inclination of their orbits as $\pi/2$. The distribution of orbital periods for the mock sample is adopted from those of the KEBs. 
The joint distribution of periods and $\Delta \rm RV_{\rm max}$ for the mock eclipsing binaries is shown in Fig. \ref{fig:apdis_keb}. 
The box in the figure marks out the detectable stars with periods in the range of 0.1 day - 5 years and RV amplitude higher than 10 $\rm km \ s^{-1}$. 
About 60\% of the eclipsing binaries are detected in the simulation. 
The detection rate will be reduced to 44\% given the limitation of periods of 0.5 day - 5 years. 
The simulation provides an explanation for the detection ratio $\sim$50\% of KEBs by LAMOST. 

\begin{figure}[h!]
\centering\includegraphics[width=0.85\columnwidth]{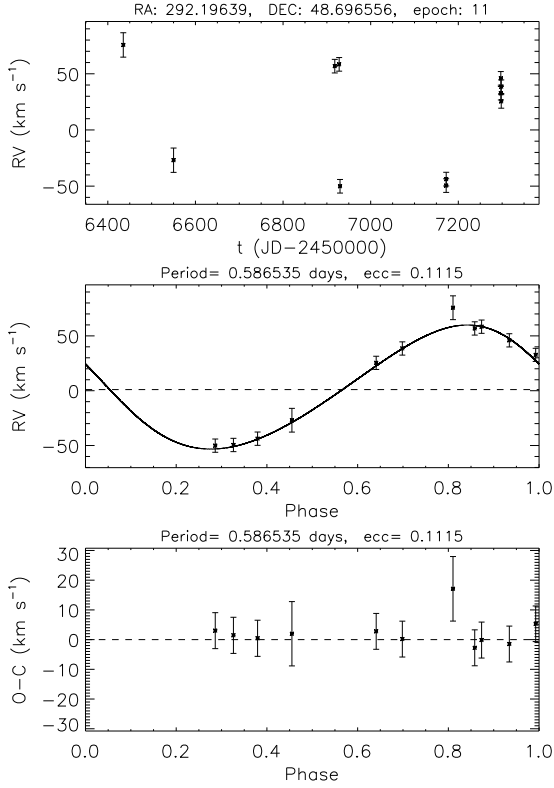}
\caption{The comparison of RVs between observations and fittings for KIC 11084782. The top panel shows the observed RVs against time. The observed (dots with error bars) and fitted (solid line) RVs against phases are plotted in the middle panel, and the residuals (O-C) are plotted in the bottom panel. \label{fig:kic1}}
\end{figure}

\begin{figure}[h!]
\centering\includegraphics[width=0.87\columnwidth]{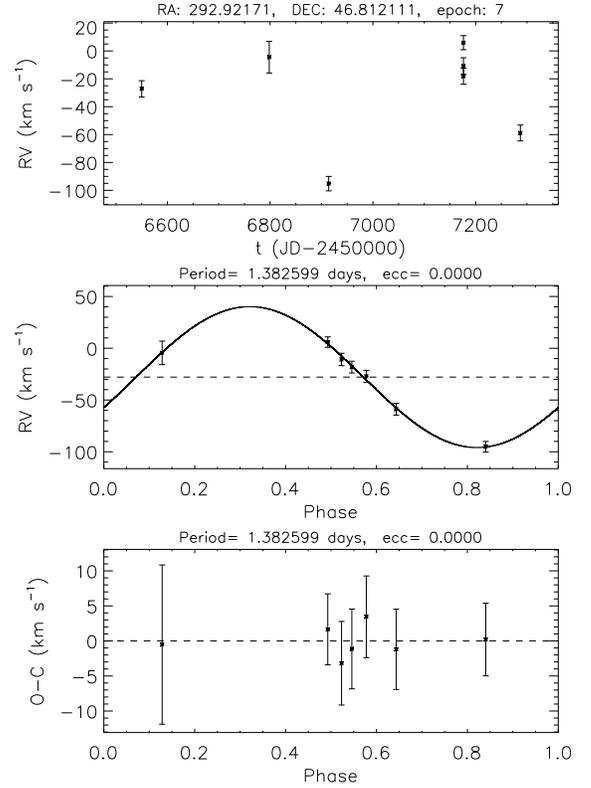}
\caption{Same as Fig. \ref{fig:kic1} but for KIC 9953894.\label{fig:kic2}}
\end{figure}

KEBs such as  KIC 11084782 and KIC 9953894 have been observed in 11 and 7 epochs by LAMOST, respectively. 
Their RV time series are plotted in the top panels of Figs. \ref{fig:kic1} and \ref{fig:kic2}. 
Given the orbital period measured with $\it Kepler$, we could fit the RVs of the binary system accurately with {\tt rvfit}. 
The {\tt rvfit} method fits RVs of stellar binaries and exoplanets using an adaptive simulated annealing (ASA) global minimization method, which quickly converges to a global solution minimum without the need to provide preliminary parameter values. The efficiency and reliability have been verified by \citet{2015ascl.soft05020I,2015PASP..127..567I}. 
As shown in the middle panels of Figs. \ref{fig:kic1} and \ref{fig:kic2}, the observed and fitted RVs against phases are presented. The residuals (O-C) are plotted in the bottom panels of the figures. 
The RVs from spectroscopic observations together with periods from photometric observations could constrain the orbital parameters well. 

\subsection{Cross-match with GDR2 variables}\label{subsec:gdr2}
\begin{figure}
\centering\includegraphics[width=0.85\columnwidth]{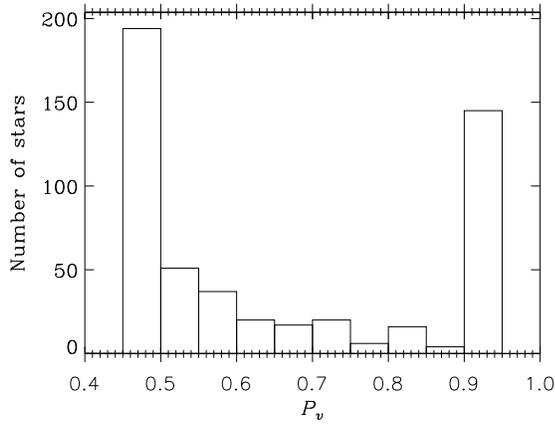}
\caption{$P_{v}$ distribution of common stars between  $\it Gaia$ DR2 (GDR2) variables and multi-epoch observed LAMOST targets.\label{fig:varipv}}
\end{figure}

\begin{figure}
\centering\includegraphics[width=0.85\columnwidth]{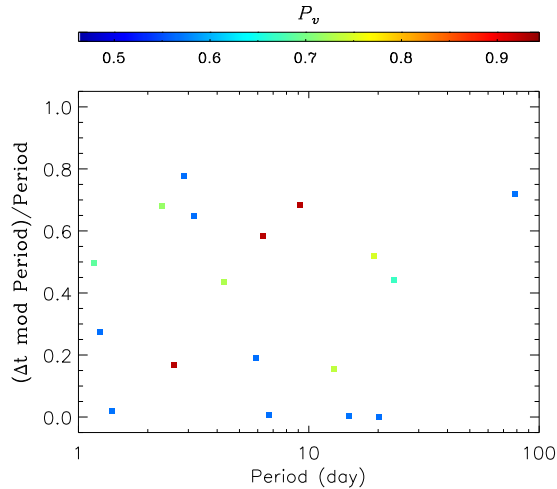}
\caption{$P_{v}$ on the Period-$\Delta \rm t$ diagram for Cepheids. The colors of points denote the probabilities of RV variable stars. }\label{fig:cepheidpv}
\end{figure}

\begin{figure}
\centering\includegraphics[width=0.85\columnwidth]{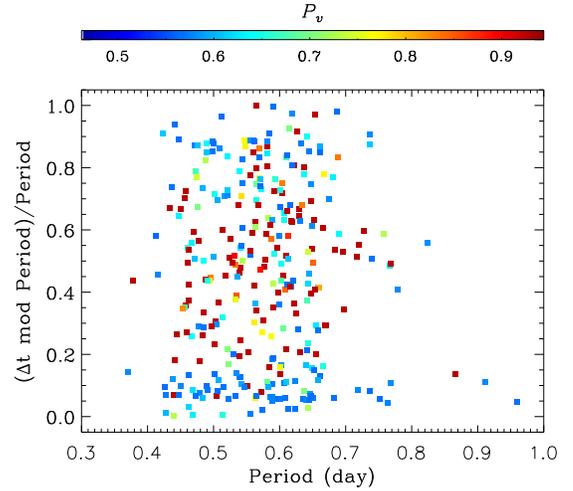}
\caption{Same as Fig. \ref{fig:cepheidpv}, but for RR Lyrae.\label{fig:rrpv}}
\end{figure}

\begin{figure}
\centering\includegraphics[width=0.85\columnwidth]{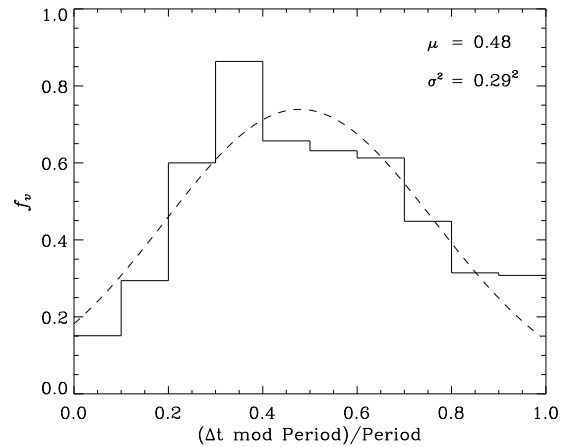}
\caption{The distribution of detection rate $f_{v}$ against the $(\Delta t \mod \rm Period)/\rm Period$ for the common RR Lyrae between GDR2 variables and LAMOST duplicated targets. The solid and dashed lines denote the calculated and Gaussian-fitted values of $f_{v}$, respectively. The mean and variance of the Gaussian distribution are shown in the figure.  \label{fig:rrpv3}}
\end{figure}

Since some stars exhibit RV variations due to periodic contraction and expansion they will, absent further characterization, contaminate the catalog of binary candidates. 
We cross-match the variable star candidates with $\it Gaia$ DR2 (GDR2) variables including Cepheids, RR Lyrae,  long-period variables (LPV) and short-period variables (SPV) \citep{2019A&A...622A..60C,2018A&A...618A..58M,2018A&A...620A.197R}. 
The distribution of $P_{v}$ for the common stars is presented in Fig. \ref{fig:varipv}. 
One hundred and ninety-eight variable stars from the 498 common sources are detected ($P_{v} >$ 0.6) with LAMOST. 
The common sources include 19 Cepheids, 442 RR Lyrae, 34 LPV and 3 SPV detected with $\it Gaia$. 
Among them, 10 Cepheids, 179 RR Lyrae, 4 LPV and 0 SPV are identified as RV variables in our database.  
The true positive rate of the catalog is about 39\% for these intrinsic variables. This value is different than that of binary systems because of the different period distributions between intrinsic and extrinsic variables.
The Period-$\Delta \rm t$ diagrams for these common Cepheids and RR Lyrae are presented in Figs. \ref{fig:cepheidpv} and \ref{fig:rrpv}, respectively. 
Their periods are provided by $\it Gaia$ variable catalogs, while the $\Delta \rm t$ are from LAMOST observations. 
From the figures, we can see that the detection rates are related to sampling characteristics of observations as well as stellar periods.  
Figure \ref{fig:rrpv3} quantifies the distribution of detection rate $f_{v}$ against the $(\Delta t \mod \rm Period)/\rm Period$ for the common RR Lyrae between GDR2 variables and LAMOST multi-epoch targets. A Gaussian curve of the $f_{v}$ with a mean of 0.48 and variance of $0.29^2$ illustrates the detection rate depends on sampling characteristics of observations and stellar periods. 

Meanwhile, we cross-match the LAMOST multi-epoch targets with the catalogue of radial velocity standard stars from GDR2 \citep{2018A&A...616A...7S}. 
None of the 7 common stars were identified as RV variables in our catalog. 

\begin{figure*}
\centering\includegraphics[width=2\columnwidth]{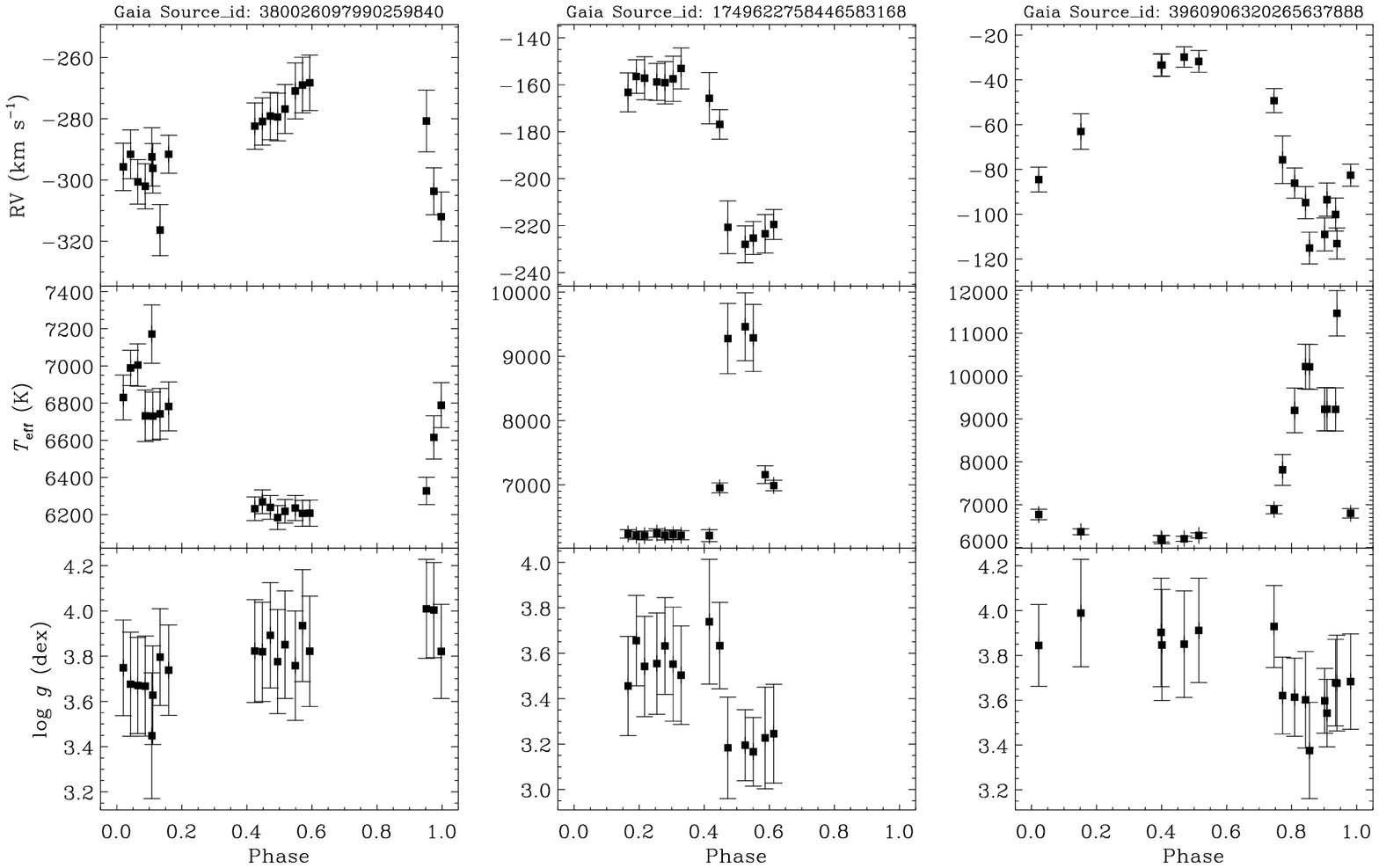}
\caption{Phased variations of RV, $T_{\rm eff}$ and $\log g$ for examples of the RR Lyrae stars.\label{fig:rrf}}
\end{figure*}

As examples indicated in Fig. \ref{fig:rrf}, the phased variations of RV, $T_{\rm eff}$ and $\log g$ for RR Lyrae are presented. 
Their periods are measured with $\it Gaia$ and stellar parameters and RVs are derived from LAMOST spectra. 
Since the pulsation of an RR Lyra, its $\log g$ varies with the radius directly, meanwhile, 
its $T_{\rm eff}$ decreases and increases with the contraction and expansion of the star, respectively. 
The variations of stellar parameters and RVs could be detected through the LAMOST observations. 
A detailed analysis of RR Lyrae observed with LAMOST is presented in \citet{2020ApJS..247...68L} and interested readers please refer to the paper. 

\begin{deluxetable*}{ccccccccc}
\tablecaption{Common sources of LAMOST and $\it Gaia$ SPV. \label{tab:spv}}
\tablecolumns{8}
\tablewidth{0pt}
\tablehead{No. &   $Ra$   & $Dec$ & epochs & $\Delta \rm RV_{max}$ ($\rm km \ s^{-1}$) & $\overline{\sigma_{\rm RV}}$ ($\rm km \ s^{-1}$) &  $P_{v}$ & $\rm Amplitude$ (mag) & Frequency ($\rm d^{-1}$)  }
\startdata
           1 &     119.46654 &       20.091425 &           2 &      4.88     &         5.79 &     0.46 &  0.22 & 9.30  \\
           2 &     205.83937 &       20.921980 &           2 &      22.83   &       14.85 &     0.48 & 0.68  & 43.20 \\
           3 &     200.50029 &       18.452419 &           2 &      35.28   &       11.54 &     0.58 & 0.94  & 2.03    \\
\enddata
\end{deluxetable*}

Note that the LAMOST is not adequate to detect short-period RV variables with periods shorter than two hours based on Nyquist's theorem, especially for extreme short-period ones, since the typical exposure time of LAMOST is in the order of an hour. 
We list the 3 common sources between $\it Gaia$ SPV and multi-epoch observed LAMOST targets in Table \ref{tab:spv}. 
From the table, we can see that low-period (high-frequency) SPV could not be detected as variables with LAMOST. 
It demonstrates that the probability of detection is related to the period (or frequency) of the target. 

\subsection{Cross-match with VSX}
In order to investigate our catalog further, we cross-match the catalog with other variable stars published in literatures. 
The VSX is a comprehensive relational database of known and suspected variable stars gathered from a variety of respected published sources \citep[][]{2006SASS...25...47W}. 
About 600,000 variable stars are collected and about three-fourths of them are provided with types and periods in VSX. There are 10,557 shared sources between VSX and LAMOST duplicated targets. Among them, 3,044 stars are detected as RV variables in our catalog. The types of the detected stars include binary stars and pulsating stars. The comprehensive detection rate of VSX is about 29\% by LAMOST. 

\subsection{Cross-match with GCVS}
The GCVS is another catalog of variable stars. The GCVS 5.1 version contains data for 53,626 individual variable stars discovered and named as variable stars by 2017 and located mainly in the Galaxy \citep[Version 5.1,][]{2017ARep...61...80S}. An assemblage of 33,264 variables is provided with types and periods in GCVS 5.1. 
Among 924 common sources between GCVS 5.1 and LAMOST multi-epoch sources, 453 stars are recognized as RV variables in our catalog. The comprehensive detection rate of GCVS is about 49\% by LAMOST.  

\subsection{Cross-match with ASAS-SN}
The All-Sky Automated Survey for SuperNovae (ASAS-SN) scans the extragalactic sky visible from Hawaii roughly once every five nights in the V-band \citep{2014ApJ...788...48S}. Catalogs of variable stars based on ASAS-SN have been released by \citet{2018MNRAS.477.3145J,2019MNRAS.486.1907J,2019MNRAS.485..961J}. These catalogs collect 542,526 variable stars including 334,095 supplied with types and periods. 
There are 5,113 common sources between the ASAS-SN variable catalogs and LAMOST multi-epoch targets. Among them, 2,011 stars are recognized as RV variables in our catalog. 
The comprehensive detection rate of ASAS-SN variables is about 39\% by LAMOST. 

\subsection{Characteristics of the catalog}
\label{classification}
A summary of the numbers of common sources between the published catalogs and LAMOST multi-epoch targets are listed in Table \ref{tab:cs}. 
Note that some variable stars are identified repeatedly in different published catalogs. There are 11,035 common sources between LAMOST multi-epoch targets and the referred variable catalogs such as KEBs, GDR2 variables, VSX, GCVS, and ASAS-SN variable catalogs. 3,163 common sources are detected as RV variables in our catalog. The detection rate of our catalog is 29\% for the variables published in the referred catalogs. 

\begin{deluxetable*}{lccc}
\tablecaption{Numbers of common sources and detected sources by LAMOST. \label{tab:cs}}
\tablecolumns{4}
\tablewidth{0pt}
\tablehead{Catalog &  Common sources with LAMOST & Detected by LAMOST & Detection rate  }
\startdata
     KEBs           &        520       &      255     &       0.49     \\
     GDR2         &        495        &      190     &        0.38     \\
     VSX            &   10,557       &   3,044      &        0.29     \\
     GCVS         &        924       &      453      &        0.49     \\
     ASAS-SN   &     5,113       &     2,011     &        0.39     \\
\enddata
\end{deluxetable*}

Variable stars fall into two categories: intrinsic and extrinsic variables. 
Binaries belonging to extrinsic variables and pulsating stars from intrinsic ones could be detected through variations of RVs based on the LAMOST's capability. 
There are 80,702 stars detected as RV variables among the 818,136 stars with multiple epochs by LAMOST. 
As discussed in Sections \ref{subsec:keb} and \ref{subsec:gdr2}, not only binaries are included in the catalog, but also some intrinsic variables such as RR Lyrae and Cepheids. 
According to the CSP simulation in Section \ref{method1}, about 8\% of the sample are detected as binaries with a purity of 80\%, which implies that 6.4\% of the LAMOST targets with multiple epochs are binaries and 1.6\% ($\sim$13,000) are pollution by single stars given the LAMOST multi-epoch targets consist of single and binary stars. 
However, the 80,702 detected stars dominate about 10\% of the LAMOST targets with multiple epochs, which is higher than the detection rate about 8\% in the CSP simulation. 
Consequently, the others (15,251 stars) in the catalog, dominating approximately 2\% of the LAMOST multi-epoch sources, probably consist of some intrinsic variables and pollution by single stars. Applying the curve of pulsating star fractions against $T_{\rm eff}$ \citep[see Fig. 11 in][]{2019MNRAS.485.2380M} in the LAMOST targets with multiple epochs, the number of pulsating stars covered by LAMOST is expected to be approximately 20,000. However, only pulsating stars with period and RV amplitude in a specified range could be detected by LAMOST.  Assuming a typical detection rate 30\% of the pulsators, the number of detected pulsating stars in our catalog is approximately 6,000. 
Thus, the 15,251 stars are mainly constructed with binary stars and pulsating stars, probably. 
Therefore, the catalog consists of $\sim$62,000 binaries (77\%), $\sim$ 6,000 pulsating stars (7\%) and a pollution by $\sim$13,000 single stars (16\%). 

\begin{figure}[ht!]
\centering\includegraphics[width=0.90\columnwidth]{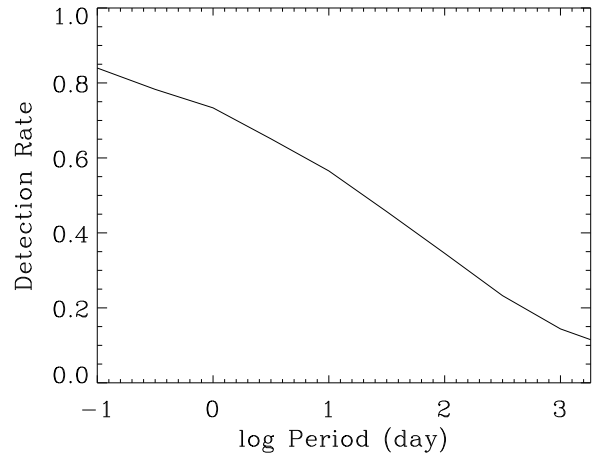}
\caption{The detection rate verses orbital periods for the BSP simulation.\label{fig:pdr}}
\end{figure}

\begin{figure}[ht!]
\centering\includegraphics[width=0.90\columnwidth]{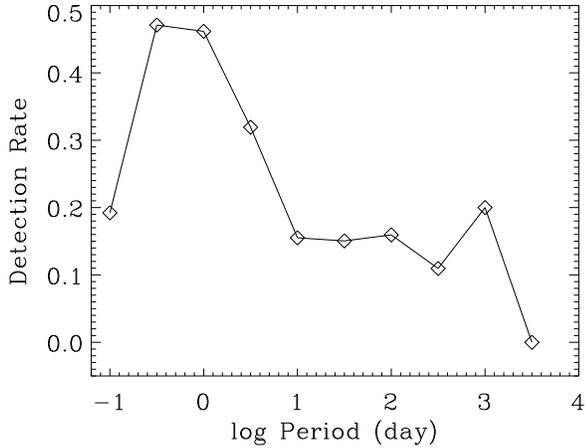}
\caption{The detection rate of RV variables by LAMOST for the common sources between LAMOST multi-epoch targets and published catalogs. \label{fig:pdra}}
\end{figure}

Based on the BSP simulation in Section \ref{method1}, the detection rates of binaries against their periods are presented in Fig. \ref{fig:pdr}.  The detection rates drop exponentially with the increasing of periods. 
Figure \ref{fig:pdra} displays the detection rate of common sources between LAMOST repeated targets and the published catalogs mentioned before. 
The classifications of the shared stars through cross-identifications with the previous catalogs are listed in Table \ref{tab:cat}. 
The distribution of detection rate indicates that the method adopted in this work based on $\Delta \rm RV_{\rm max}$ by LAMOST is sensitive to short-period RV variable stars such as short-period binaries and RR Lyrae.  All the same, various types of variable stars appear in our catalog. However, most of the variables collected in our catalog, so far, are not able to be classified based on LAMOST spectra or data from other surveys. 

\section{Conclusions and Discussions} \label{sec:summary}
We analyze the probabilities of being RV variable stars based on the duplicated observations for LAMOST DR4 targets. 
A catalog of 80,702 RV variable star candidates is constructed. 
The false positive rate of the catalog is about 3\% based on the LAMOST ability. 
The purity of the catalog is estimated to be better than $\sim$80\% through simulation and cross-identifications.
Both intrinsic and extrinsic variable stars are collected in the catalog. 
It consists of 77\% binary systems and 7\% pulsating stars as well as 16\% pollution by single stars. 
The catalog is a powerful database of RV variable candidates, which could be taken as an input source for RV variable surveys. 

Since some intrinsic variables present variability of RV, the catalog is blended with pulsating stars such as Cepheids, RR Lyrae, LPVs and SPVs. 
The cross-identifications and classifications are carried out by matching with $\it Kepler$ Eclipsing Binaries, GDR2 variables, VSX, GCVS, and ASAS-SN variables.  A number of 3,138 stars in our catalog are classified. Although recognized as RV variables, most of the variable stars in the catalog are not classified based on the LAMOST data or other surveys. 
The efficiency of the method adopted in this work relies on not only sampling frequency of observations but also periods and amplitudes of variable stars.

The key foundation of this work is the accuracy of RVs and their uncertainties. Fortunately, over-estimating uncertainties will not affect the accuracy of identifying RV variables or their candidates, although some of them would be left out. 
In future work, we will make use of spectral and photometric data from LAMOST and other surveys to classify the catalog of RV variable stars as a follow-up to this work. The spectra of classified stars would be adopted as training set to recognize spectra of unclassified RV variables based on machine-learning method, probably. Meanwhile, the common sources between the RV variables and X-ray sources will provide more clues of binary interactions. 

\acknowledgments
This work has made use of data products from the Guoshoujing Telescope (the Large Sky Area Multi-Object Fibre Spectroscopic Telescope, LAMOST). 
LAMOST is a National Major Scientific Project built by the Chinese Academy of Sciences. Funding for the project has been provided by the National Development and Reform Commission. LAMOST is operated and managed by the National Astronomical Observatories, Chinese Academy of Sciences.

This work is partially supported by National Natural Science Foundation of China 11803030, 11443006, 11773005, 11803029, National Key Basic Research Program of China 2014CB845700, China Postdoctoral Science Foundation 2016M600850, Science \& Technology Department of Yunnan Province - Yunnan University Joint Funding (2019FY003005) and Joint Research Fund in Astronomy U1531244 and U1631236. 
The LAMOST FELLOWSHIP is supported by Special Funding for Advanced Users, budgeted and administrated by Center for Astronomical Mega-Science, Chinese Academy of Sciences (CAMS).

We would like to thank the anonymous referee for valuable comments which improved the manuscript.

\begin{longrotatetable}
\begin{deluxetable*}{cccccccccccccc}
\tablecaption{Catalog of RV variable star candidates.  \\
The $Ra$ and $Dec$ of the stars are listed in columns 2--3. 
Number of epochs and time duration of observations for each star are shown in columns 4--5. 
The maximum variation of RV and the weighted error are listed in columns 6--7. 
SNRs and time of exposures responding to the maximum and minimum RVs are listed in columns 8--11. 
The probability of being a RV variable star is provided in the last column. 
LAMOST unique spectral ID, SNR, time for each exposure, stellar parameters and RVs together with their errors of each epoch see a detailed and inclusive version of the catalog online.
\label{tab:cat}}
\tablecolumns{14}
\tablewidth{900pt}
\tabletypesize{\scriptsize}
\tablehead{
\colhead{No.} & {  $Ra$  } & $Dec$ & epochs & Time Duration  &  $\Delta \rm RV_{max}$ & $\overline{\sigma_{\rm RV}}$ &  $\rm SNR_{\rm RV_{max}}$  &  
 $\rm SNR_{\rm RV_{min}}$ &  $\rm t_{\rm RV_{max}}$ &  $\rm t_{\rm RV_{min}}$ & $P_{v}$  & Classification & Notes\tablenotemark{a}  \\ }
\startdata
1  &   0.0170327    & 56.0176468    &  2    &    707.0498657  &       40.3  &    7.5  &     11  &    21   &    2456968.093  &   2456261.043  &   0.95 &  &  \\
2  &   0.0204036    & 61.5655899    &  2    &    414.8555298  &       27.2  &    8.4  &     77  &    39   &    2457324.099  &   2456909.243  &   0.62 &  &  \\
3  &   0.0236430    & 60.1553001    &  2    &    412.9315491  &       37.2  &  11.4  &     12  &    25   &    2457322.097  &   2456909.166  &   0.63 &  &  \\
4  &   0.0246262    & 62.7630920    &  2    &    412.8926392  &       18.1  &    5.4  &     16  &    32   &    2457322.097  &   2456909.204  &   0.65 &  &  \\
5  &   0.0312140    & 35.5037842    &  2    &  1029.2077637  &       15.2  &    4.4  &     70  &   183  &    2456262.009  &   2457291.217  &   0.65 &  &  \\
6  &   0.0470053    & 61.6638145    &  2    &    412.9263611  &       17.2  &    4.2  &     95  &    27   &    2456909.204  &   2457322.131  &   0.76 &  &  \\
7  &   0.0742439    & 60.3994408    &  2    &    412.9653015  &       41.7  &    7.1  &     16  &   124  &    2457322.131  &   2456909.166  &   0.95 &  &  \\
8  &   0.0804520    & 37.1159744    &  2    &    674.0565796  &       25.5  &    4.3  &     28  &    89   &    2456262.037  &   2456936.094  &   0.95 &  &  \\
9  &   0.0865299    & 36.5521774    &  2    &    748.9668579  &       16.1  &    4.0  &     19  &    96   &    2456261.981  &   2457010.948  &   0.76 &  &  \\
10  &  0.0902442   & 55.8368454    &  2    &      28.9211788  &       31.0  &    7.3  &     53  &    91   &    2456968.049  &   2456996.970  &   0.82 &  &  \\
          ... & ...  & ... & ...  &  ... & ...  &  ...  & ...  & ...  & ...  & ...  & ...  &  ...  & \\
\enddata
\tablenotetext{a}{The Notes label marks the common sources between LAMOST and other surveys.}
\end{deluxetable*}
\end{longrotatetable}

\bibliography{example}



\end{document}